\documentclass[
    ,final            
  ]
  {aipproc}

\layoutstyle{6x9}

\begin{document}

\title{Relativistic Shock Acceleration: A Hartree-Fock Approach}

\classification{96.50.Pw }
\keywords      {Particle Acceleration, Relativistic Shocks}
\author{Paul Dempsey}{
  address={School of Cosmic Physics, Dublin Institute for Advanced Studies, 31 Fitzwilliam Pl., Dublin 2, Ireland. pdempsey@cp.dias.ie},
}
\author{John Kirk}{
  address={Max-Planck-Institut f\"{u}r Kernphysik,  Heidelberg 69029, Germany. john.kirk@mpi-hd.mpg.de},
}

\begin{abstract}
We examine the problem of particle acceleration at a relativistic shocks assuming pitch-angle scattering and 
using a Hartree-Fock method to approximate the associated eigenfunctions. This
leads to a simple transcendental equation determining the power-law index, $s$, 
given the up and downstream velocities. We compare our results with accurate numerical 
solutions obtained using the eigenfunction method. In
addition to the power-law index this method yields the angular and
spatial distributions upstream of the shock. 
\end{abstract}

\maketitle


Particle acceleration at relativistic shocks via the first-order Fermi process has been the subject of a great deal of analysis since the late eighties. Initial calculations using semi-analytical \cite{KS1987, HD1988, KGGA2000} and Monte Carlo \cite{AGKG2001} methods based on pitch-angle scattering
suggested a ``universal'' power law index for ultra-relativistic shocks of $s=4.22$. However, recent simulations that compute particle trajectories in prescribed random fields present a more complicated picture
\cite{LPR2006,NO2006,NOP2006,PLM2008}. Both approaches involve essentially arbitrary assumptions about the particle transport properties, and the only realistic prospect of making progress lies in particle-in-cell simulations. Recently, these have produced the first evidence that relativistic shocks can self-consistently accelerate particles \cite{Spitkovsky2008}, and the results appear to favour the pitch-angle scattering approach, which motivates us to present here an improved approximation 
scheme based on \cite{KGGA2000}. 
A simple formula relating the power law index $s$ to the shock velocities that provides a reasonable fit to the numerical results has already been advanced by \citet{KW2005}. But, as these authors note, its derivation is incomplete. In this contribution we present a simple analytical approximation scheme, based on the variational method familiar from Hartree-Fock computations of atomic eigenfunctions. The results give both the angular distribution and the spectal index and are in good agreement with accurate numerical evaluations. 



The particle transport equation satisfied both upstream and downstream is
\begin{eqnarray}\label{transport_urel}
\Gamma\, (\beta+\mu){\partial f \over \partial z} ={\partial \over \partial \mu}\left(D(1-\mu^2){\partial f \over \partial \mu}\right)
\end{eqnarray} 
where $\beta$ is the speed of the fluid with respect to the shock front (in which the distribution is assumed stationary) and $\Gamma=\left(1-\beta^2\right)^{-1/2}$.
We look for solutions of the form $f(p,\mu,z)=\sum_if_ip^{-s}Q_i(\mu)\exp ( \Lambda_i z/\Gamma)$ with eigenfunction-eigenvalue pairs satisfying:
\begin{eqnarray}
 \frac{\partial}{\partial\mu}\left(D(1-\mu^2)\frac{\partial{Q_i}}{\partial{\mu}}\right)= \Lambda_i(\beta+\mu)Q_i
\label{eigenfunctioneq}
\end{eqnarray}
It has previously been shown that the upstream angular distribution can be approximated very accurately by the 
eigenfunction $Q_1$ of smallest positive eigenvalue, which, in the ultra-relativistic case $\Gamma\gg1$, is approximately 
$Q_1\sim\exp(-(1+\mu)/(1-\beta))$ \cite{KGGA2000}. The value of the spectral index follows when the boundary condition far downstream 
is applied to this function, after is has been transformed into the downstream frame. This requires knowledge of $Q_1$ at the downstream speed $\beta=\beta_{\rm d}$, which is not ultra-relativistic. 

Using this information we chose a one parameter trial function for the upstream eigenfunction of the form
$Q_1=\exp(a_u\mu)$.
$a_u$ is determined by the requirement that $Q_1$ be orthogonal to the isotropic eigenfunction of zero eigenvalue, $Q=\,$constant:
$\exp(2a_u)=({a_u(\beta-1)-1})/({a_u(\beta+1)-1})$.
The corresponding eigenvalue is found by projecting Eq.~(\ref{eigenfunctioneq}) onto $Q_1$, giving
\begin{eqnarray}
 \Lambda_1=D\frac{-2a_u^2\left(e^{4a_u}+1\right)+a_u\left(e^{4a_u}-1\right)}{e^{4a_u}\left(2a_u\left(\beta+1\right)-1\right)-\left(2a_u\left(\beta-1\right)-1\right)},
\end{eqnarray}
which differs from the numerically calculated eigenfunction by less than $0.2\%$ (this occurs near $\beta=1/3$). 
For the downstream eigenfunction, choose a two-parameter polynomial trial function:
\begin{eqnarray}
 Q_1=1+a_{d1}\mu +a_{d2}\mu^2.
\end{eqnarray}
As in the Hartree-Fock method, the parameters are evaluated by minimising the eigenvalue. A minor subtlety arises here, because Eq.~(\ref{eigenfunctioneq}) has a family of eigenfunctions of negative sign. However, these are easily excluded in the minimising prodedure. 
Enforcing the orthogonality condition and minimising the eigenvalue we obtain 
\begin{eqnarray}
 a_{d2}=-\left(\frac{a_{d1}+3\beta}{\beta}\right)
\qquad a_{d1}=-\beta\frac{-13+10\beta^2-\sqrt{25-20\beta^2+100\beta^4}}{2(-3+5\beta^2)}.
\end{eqnarray}
This trial function is only valid for $\beta<\sqrt{3/5}$. The corresponding eigenvalue is
\begin{eqnarray}
 \Lambda_1 = -D \frac{10 a_{d1}^2 + 8a_{d2}^2}{10a_{d1} +6a_{d1}a_{d2} + \beta\left(15 + 5a_{d1}^2+3a_{d2}^2+10a_{d2}\right)},
\end{eqnarray}
which again closely agrees with the numerically calculated value.


The spectral index $s$ is the solution of the integral equation \cite{KGGA2000}
\begin{eqnarray}
 \int_{-1}^{1} d\mu_u(\beta_u+\mu_u)Q_1^d(\mu_d) Q_1^{u}(\mu_u) (1+\beta_{\rm rel}\mu_u)^{s-3}=0,
\end{eqnarray}
where the indices $u$ and $d$ refer to quantities upstream and downstream respectively.\\
With the substitution $t=-a_u(1+\beta_{\rm rel}\mu_u)/\beta_{\rm rel}$, the integral reduces to
\begin{eqnarray}
 \int_{t_1}^{t_2} t^{s-5}\left(P_0 +P_1t+P_2t^2+P_3t^3\right)e^{-t}\;dt=0   
\end{eqnarray}
where $
 \qquad t_1= -a_u(1-\beta_{\rm rel})/\beta_{\rm rel}; \qquad t_2 = -a_u(1+\beta_{\rm rel})/\beta_{\rm rel}
$ and
\begin{eqnarray}
 P_0&=&a_{d2}\frac{1-\beta_{\rm rel}\beta_u}{\beta_{\rm rel}}\left(\frac{1-\beta_{\rm rel}^2}{\beta_{\rm rel}}\right)^2 \\
P_1&=&\left(
\left(a_{d1}\beta_{\rm rel} + 2a_{d2}\right)\frac{1-\beta_{\rm rel}^2}{\beta_{\rm rel}}\frac{1-\beta_{\rm rel}\beta_u}{\beta_{\rm rel}} 
+ a_{d2}\left(\frac{1-\beta_{\rm rel}^2}{\beta_{\rm rel}}\right)^2 \right)/a_{u}  \\
P_2&=&\left(\left(a_{d1}\beta_{\rm rel} + 2a_{d2}\right)\frac{1-\beta_{\rm rel}^2}{\beta_{\rm rel}}
+ \frac{1-\beta_{\rm rel}\beta_u}{\beta_{\rm rel}}\left(\beta_{\rm rel}^2 + a_{d1}\beta_{\rm rel} + a_{d2}\right)\right)/a_{u}^2 \\
P_3&=&\left(\beta_{\rm rel}^2 + a_{d1}\beta_{\rm rel} + a_{d2}\right)/a_{u}^3.
\end{eqnarray}
Therefore, $s$ satisfies  
\begin{eqnarray}
 P_3\left(\Gamma(s-1,t_1)-\Gamma(s-1,t_2)\right)+
P_2\left(\Gamma(s-2,t_1)-\Gamma(s-2,t_2)\right)+\nonumber\\
P_1\left(\Gamma(s-3,t_1)-\Gamma(s-3,t_2)\right)+
P_0\left(\Gamma(s-4,t_1)-\Gamma(s-4,t_2)\right)=0,
\end{eqnarray}
where $\Gamma(x,t)$ is the incomplete gamma function. This transcendental equation can be solved easily with any root finding algorithm. 


\begin{figure}  
\includegraphics[height=.3\textheight]{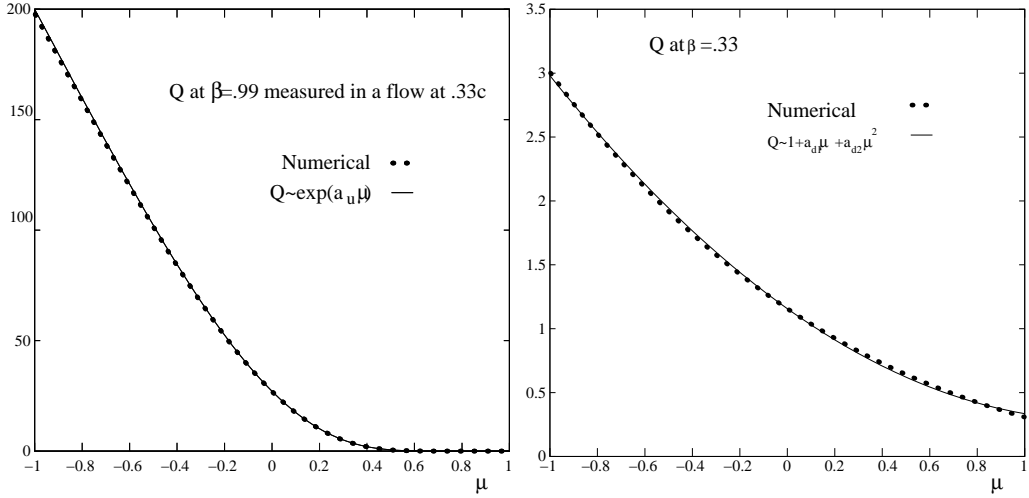}
\label{fig:combofirst}
\caption{Left: the numerically obtained eigenfunction compared to the single-parameter approximation to $Q_1$ for $\beta=.99$. 
$\mu$, is the 
angle between the shock normal and the particle momentum measured in a frame of reference (effectively the downstream frame) 
moving at $\beta=.33$. Right: the numerically obtained eigenfunction compared to $Q_1\sim 1+a_{d1}\mu+a_{d2}\mu^2 $ for $\beta=.33$.}

\end{figure}

In figure {\ref{fig:combofirst}} we compare the eigenfunction $Q_1$ obtained numerically using the Pr\"ufer transformation \cite{HD1988,KGGA2000} with our Hartree-Fock approximation for two flow speeds typical of upstream and downstream conditions. Agreement is very good for both $\beta=.99$ and $\beta=.33$. In the $\beta=.33$ case the approximate eigenvalue differs from the eigenvalue obtained numerically by about $0.2\%$. This inaccuracy has a small effect on the value of the ultra-relativistic spectral index obtained, $s=4.23$, about $0.23\%$ higher than the accurate result of $s=4.22$, obtained using numerical evaluation of the eigenvalues and expanding the distribution function to higher order than $Q_1$ \cite{KGGA2000}. 

\begin{figure}
\includegraphics[height=.3\textheight]{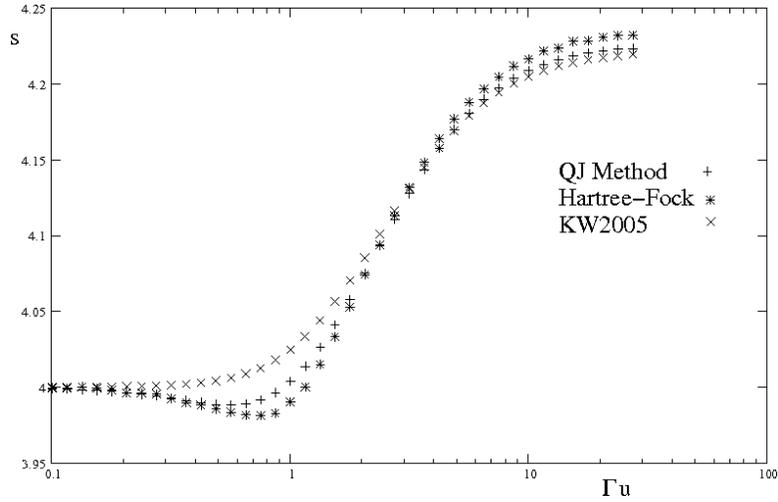}
\label{fig:logsJS}
\caption{The spectral index obtained for the J\"uttner-Synge equation of state via the $Q_J$ method, Hartree-Fock approximation and the fit $s=(3\beta_u - 2\beta_u\beta_d^2+\beta_d^3)/(\beta_u-\beta_d)$ are compared. 
}
\end{figure}

In figure {\ref{fig:logsJS}} we compare the value of $s$ obtained from the Hartree-Fock approximation with accurate numerical results and with the formula 
proposed by Keshet \& Waxman \cite{KW2005} $s=(3\beta_u - 2\beta_u\beta_d^2+\beta_d^3)/(\beta_u-\beta_d)$. In this example, 
the upstream and downstream speeds are related by the jump condition given by the 
J\"uttner-Synge equation of state. The value obtained via the fitting formula is least accurate for mildly relativistic shocks ($.5<\Gamma u<2$), where
it deviates by $\Delta s\approx 0.03$. The Hartree-Fock approximation deviates by at most $\Delta s=.015$ for all shock velocities.


\begin{theacknowledgments}
Paul Dempsey would like to thank the Irish Research Council for Science, Engineering and Technology and the Max Planck Society for supporting this work.
\end{theacknowledgments}



\bibliographystyle{aipproc}   



\end{document}